\documentclass{ws-ijmpe}
\usepackage[latin1]{inputenc}
\begin{document}

\markboth{B. Borderie et al.}{The prominent role of the heaviest fragment in multifragmentation
and phase transition for hot nuclei}

\catchline{}{}{}{}{}

\title{THE PROMINENT ROLE OF THE HEAVIEST FRAGMENT IN MULTIFRAGMENTATION
AND PHASE TRANSITION FOR HOT NUCLEI\\Indra and Aladin Collaborations}

\author{\footnotesize B. BORDERIE, E. BONNET\footnote{present 
address: GANIL, DSM/CEA-CNRS/IN2P3}, N. LE
NEINDRE\footnote{present address: LPC Caen, CNRS/IN2P3, ENSICAEN, Univ. de Caen},
S. PIANTELLI\footnote{present address: Sezione INFN, Sesto Fiorentino (Fi), Italy},\\
Ad. R. RADUTA\footnote{permanent address: NIPNE, Bucharest-M\u{a}gurele, Romania },
M. F. RIVET, E. GALICHET}

\address{Institut de Physique Nucl\'eaire, CNRS/IN2P3, Univ. Paris-Sud 11\\
91406, Orsay cedex, France,\\
borderie@ipno.in2p3.fr}

\author{F. GULMINELLI, D. MERCIER\footnote{permanent address: Institut de
Physique Nucl\'eaire, CNRS/IN2P3 et Univ. Lyon}, B. TAMAIN,
R. BOUGAULT, M. P\^ARLOG\footnote{permanent address: NIPNE,
Bucharest-M\u{a}gurele, Romania}}

\address{LPC Caen, CNRS/IN2P3, ENSICAEN, Univ. de Caen\\
14050, Caen cedex, France\\}

\author{J. D. FRANKLAND, A. CHBIHI, J. P. WIELECZKO}

\address{GANIL, DSM/CEA-CNRS/IN2P3\\
14076, Caen cedex, France\\}

\author{D. GUINET, P. LAUTESSE}

\address{Institut de Physique Nucl\'eaire, CNRS/IN2P3 et Univ. Lyon 1\\
69622, Villeurbanne cedex, France\\}

\author{F. GAGNON-MOISAN. R. ROY}

\address{Laboratoire de Physique Nucl\'eaire, Univ. Laval\\
G1K7P4, Qu\'ebec, Canada\\}

\author{M. VIGILANTE, E. ROSATO}

\address{Dip. di Scienze Fisiche e Sezione INFN, Univ. di Napoli\\
80126, Napoly, Italy\\}

\author{R. DAYRAS}

\address{IRFU/SPhN, CEA Saclay\\
91191, Gif sur Yvette, France\\}

\author{J. {\L}UKASIK}

\address{Institute of Nuclear Physics, IFJ-PAN\\
31342, Krak\'ow, Poland\\}

\maketitle

\begin{history}
\received{(received date)}
\revised{(revised date)}
\end{history}

\begin{abstract}
The role played by the heaviest fragment in partitions of multifragmenting
hot nuclei is emphasized. Its size/charge distribution (mean value, fluctuations
and shape) gives information on properties of fragmenting nuclei and on
the associated phase transition.
\end{abstract}

\section{Introduction}
Nuclear multifragmentation was predicted long ago~\cite{Boh36} and studied 
since the early 80's. The properties of fragments which are issued from the
disintegration of hot nuclei are expected to
reveal and bring information on a phase transition of the liquid-gas
type. Such a phase transition is theoretically predicted
for nuclear matter. Nuclear physicists are however dealing with finite
systems.
Following the concepts of statistical physics, a new definition of phase
transitions for such systems was recently proposed, showing that 
specific phase transition signatures could be
expected.~\cite{I46-Bor02,WCI06,Bor08} Different and coherent signals
of phase transition have indeed been evidenced.
It is only with the advent of powerful 
4$\pi$ detectors~\cite{Sou06} like INDRA~\cite{I3-Pou95} that real
advances were made. With such an array in particular the heaviest fragment
of multifragmentation partitions is well identified in charge and its
kinetic energy well measured by taking into account
pulse-height defect in silicon detectors~\cite{I14-Tab99} and effect of the
delta-rays in CsI(Tl) scintillators.~\cite{I33-Par02,I34-Par02}
This paper emphasizes the importance of
the heaviest fragment properties in relation with hot fragmenting
nuclei (excitation energy and freeze-out volume) and with the associated
phase transition of the liquid-gas type (order parameter and generic signal
for finite systems).
\section{Heaviest fragment and partitions}
 The static properties of fragments emitted
by hot nuclei formed in central (quasi-fused systems (QF) from
$^{129}Xe$+$^{nat}Sn$, 25-50 AMeV)
and semi-peripheral collisions (quasi-projectiles (QP) from 
$^{197}Au$+$^{197}Au$, 80 and 100 AMeV) have been compared 
in detail~\cite{I69-Bon08} on the excitation
energy domain 4-10 AMeV.
To do that hot nuclei showing, to a
certain extent, statistical emission features were selected.
For central collisions (QF events) one selects complete and
compact events in velocity
space (constraint of flow angle $\geq 60^{\circ}$).
For peripheral collisions (QP subevents) the selection method
applied to quasi-projectiles minimizes the contribution
of dynamical emissions by imposing a compacity of fragments in velocity space.
Excitation energies of the different hot nuclei produced are calculated
using the calorimetry procedure (see~\cite{I69-Bon08} for details).
\begin{figure}[!hbt]
\centering
\includegraphics*[scale=0.5]{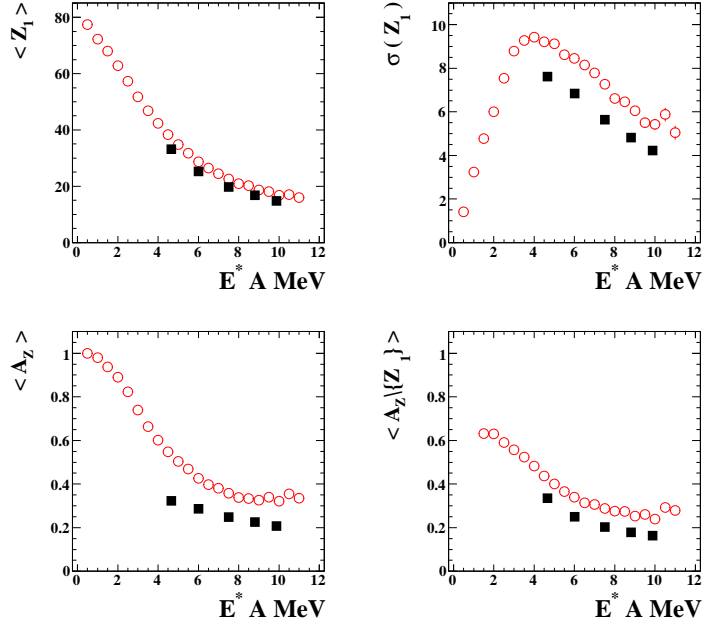}
\caption{ Full squares and open
circles stand respectively for QF and QP sources; 
Top:  average values (left) and standard deviation (right) of the charge of
the biggest fragment vs the excitation energy per nucleon. Bottom: evolution
of the charge asymmetry - with (left) and without (right) the biggest
fragment - as a function of the excitation energy per nucleon.}\label{fig1}
\end{figure}
First, it is observed that both the percentage of charge bound in fragments and
the percentage of light charged particles participating to the total charged
product multiplicity
are the same for QF and QP sources with equal excitation energy. Thus such
percentages provide a good estimate of the excitation energy of hot nuclei
which undergo multifragmentation. 

What about the size/charge of the heaviest fragment of partitions, $Z_{1}$ ?

In figure~\ref{fig1} (upper part), the evolutions, with the excitation 
energy, of its mean value and of the associated fluctuations are plotted.
The average charge of the heaviest fragment, for a given system, first 
strongly decreases with increasing excitation energy and then tends to level
off, due to the fixed lowest charge value for fragments. 
So the mean value appears as also mainly governed by excitation energy and is 
largely independent of system size and of production modes
(see~\cite{Bor08} for limitations). This 
effect was already observed in~\cite{I12-Riv98} for two QF sources
with charges in the ratio 1.5; its occurrence when comparing QF and QP
sources would indicate that their excitation energy scales do agree, within
10\%.
The fluctuations, on the contrary, exhibit sizeable differences. 
In the common energy range, the fluctuations of $Z_{1}$ decrease when 
the excitation
energy increases but they are larger for QP sources. In this latter case
they show a maximum value around 4 AMeV which is in good agreement with 
systematics reported for QP sources in~\cite{Gul06,I63-NLN07} and seems to
correspond to the center of the spinodal region as defined by the
divergences of the microcanonical heat capacity.~\cite{MDA04,T41Bon06} Differences
relative to the fluctuations of $Z_{1}$ for QF and QP sources were also
discussed in~\cite{I63-NLN07} and a possible explanation was related to different
freeze-out volumes by comparison with statistical model (SMM) calculations.
We shall see in the next section that indeed different freeze-out volumes
have also been estimated from simulations.
An overview of all information related to fragment charge partition
is obtained with the generalized  charge  asymmetry variable
calculated event by event.~\cite{I69-Bon08} To take into account
distributions of fragment multiplicities which differ for the two sources,  
the generalized asymmetry ($A_{Z}$) is introduced: 
$ A_{Z}= \sigma_{Z} / (\langle Z \rangle \sqrt{M_{frag}-1})$.
This observable evolves from 1 for asymmetric partitions to 0 for equal size
fragment partitions. 
For the one fragment events, mainly present for QP
sources, we compute the $A_{Z}$ observable by taking ``as second fragment''
the first particle 
in size hierarchy included in calorimetry. In the left bottom part of 
fig.~\ref{fig1}, 
the mean evolution with excitation energy of the generalized asymmetry
is shown. Differences are observed which well illustrate how different 
are the repartitions of $Z_{frag}$ between fragments for QF and QP 
multifragmenting sources. QP partitions are more asymmetric in the entire 
common excitation energy range. To be sure that this observation does 
not simply reflect the peculiar behaviour of the 
biggest fragment, the generalized asymmetry is re-calculated for partitions
$M_{frag}>1$, and noted 
$A_{Z} \backslash \{Z_{1}\}$, by removing $Z_{1}$ from partitions
(bottom right panel of fig.~\ref{fig1}). The difference between the 
asymmetry values for the two source types persists.
\section{Heaviest fragment and freeze-out volume}
Starting from all the available experimental information of 
selected QF
sources produced in central $^{129}$Xe+$^{nat}$Sn collisions which undergo
multifragmentation, a simulation was performed to reconstruct freeze-out
properties event by event.~\cite{I58-Pia05,I66-Pia08} 
The method requires data with a very high degree
of completeness, which is crucial for a good estimate of Coulomb energy.
The parameters of the simulation were fixed in a
consistent way including experimental partitions, kinetic properties and
the related calorimetry. The necessity of introducing a limiting temperature
related to the vanishing of level density for fragments~\cite{Koo87}
in the simulation was confirmed for all
incident energies. This naturally leads to a limitation of their excitation
energy around 3.0-3.5 AMeV as observed in.~\cite{I39-Hud03}
 \begin{figure}[htb]
\begin{center}
\includegraphics*[width=0.9\textwidth]
{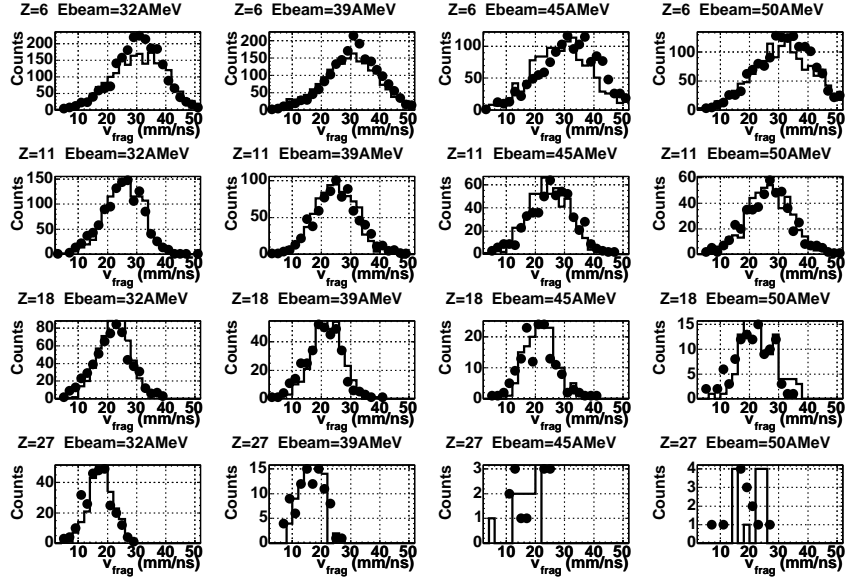}
\caption{Comparison between the experimental velocity spectra (full points) 
of fragments of a given charge and the simulated ones (histograms). 
Each row refers to a different
fragment charge: starting from the top $Z=6$, $Z=11$, $Z=18$, $Z=27$. Each
column refers to a different beam energy: starting from the left 32 AMeV,
39 AMeV, 45 AMeV and 50 AMeV. From.~\protect\cite{I66-Pia08}} \label{fig2}
\end{center}
\end{figure}
The experimental and simulated velocity spectra for
fragments of given charges (Z=6, 11, 18 and 27) are compared in 
fig.~\ref{fig2} for the
different beam energies; when
the statistics are sufficient the agreement is quite remarkable. Finally
relative velocities between fragment pairs were also compared
through reduced relative velocity correlation 
functions~\cite{Kim92,Bow95,Gro97,I57-Tab05}
(see fig.~\ref{fig3}).
\begin{figure}[htb]
\begin{center}
\includegraphics*[width=0.9\textwidth]
{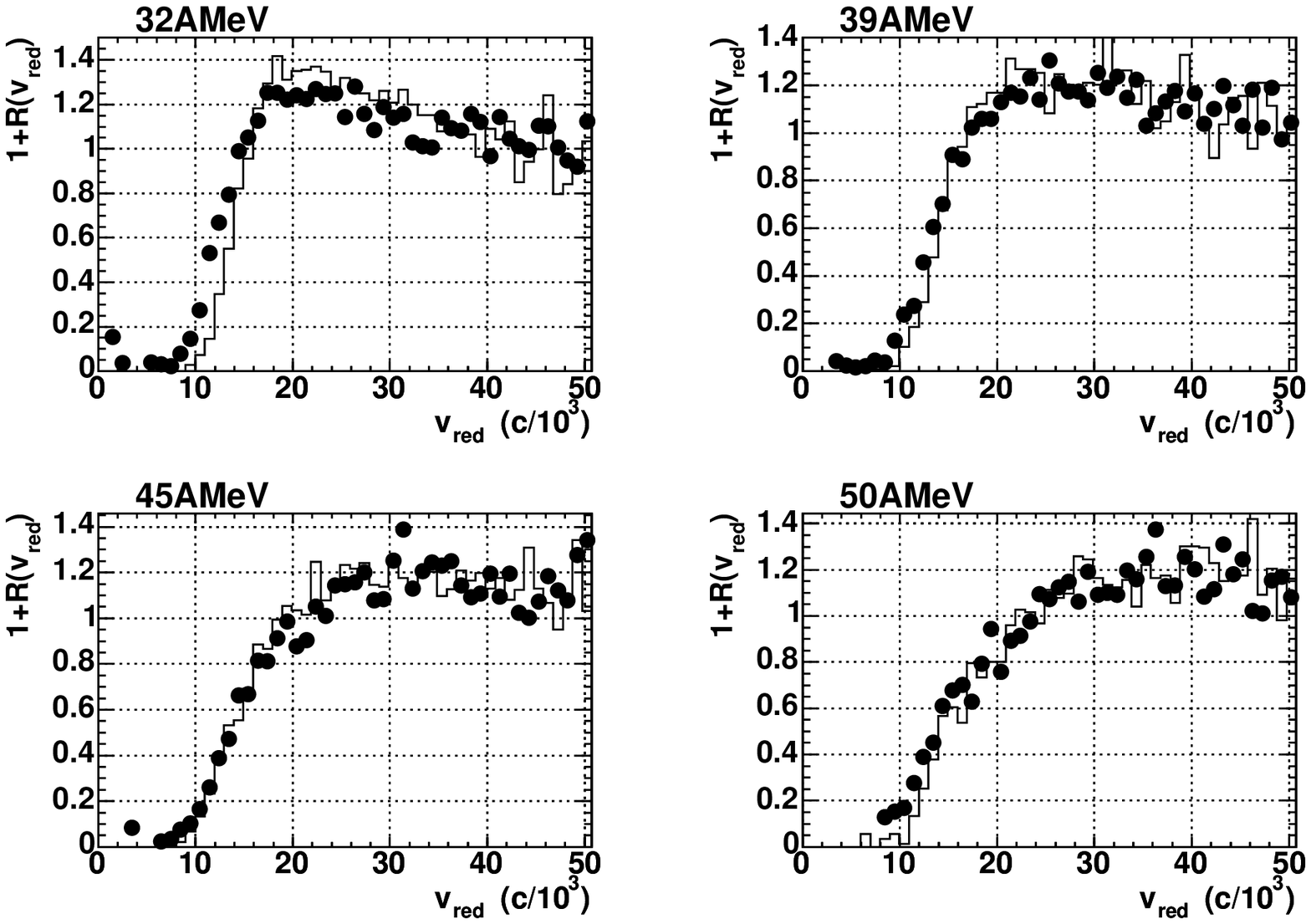}
\caption{Comparison between the experimental (full points) and simulated
(histograms) reduced relative velocity correlation functions for
all the fragments. The
reduced relative velocity between two fragments with charges $Z_i$ and $Z_j$
($Z_{i,j}>$4) is
defined as $v_{red}$=$v_{rel}$/$(Z_i+Z_j)^{1/2}$. Each panel 
refers to a different beam energy: 32 AMeV (top left),
39 AMeV (top right), 45 AMeV (bottom left) and 50 AMeV
(bottom right). From.~\protect\cite{I66-Pia08}}
\label{fig3}
\end{center}
\end{figure}
In the simulation the fragment
emission time is by definition equal to zero and correlation
functions are consequently only sensitive 
to the spatial arrangement of fragments at break-up and the
radial collective energy involved
(hole at low reduced
relative velocity), to source sizes/charges and to
excitation energy of the sources (more or less pronounced bump
at $v_{red}$= 0.02-0.03c).
Again a reasonable agreement is obtained between experimental data and
simulations, especially at 39 and 45 AMeV incident energies, which
indicates that the retained method and parameters are sufficiently relevant
to correctly describe freeze-out topologies and properties.

 The major properties of the freeze-out configurations
thus derived are the following: an important increase, 
from $\sim$20\% to $\sim$60\%, of
the percentage of particles present at freeze-out between 32 and 45-50 AMeV 
incident energies accompanied by a weak increase of the freeze-out volume 
which  tends to saturate at high excitation energy.  
Finally, to check the overall physical coherence of the developed approach,
a detailed comparison with a microcanonical statistical model (MMM) was
done. The degree of agreement, which was found acceptable, confirms the main
results and gives confidence in using those reconstructed freeze-out events
 for further studies as it is done in.~\cite{I69-Bon08}

Estimates of freeze-out volumes for QF sources produced 
in Xe+Sn collisions for incident energies between 32 and 50 AMeV
evolve from 3.9 to 5.7 $V/V_0$, where $V_0$ would correspond to 
the volume of the source at normal density.~\cite{I66-Pia08} 

\begin{figure}[htb]
\begin{minipage}[c]{.45\textwidth}
\centering
\includegraphics[width=1.15\textwidth]
{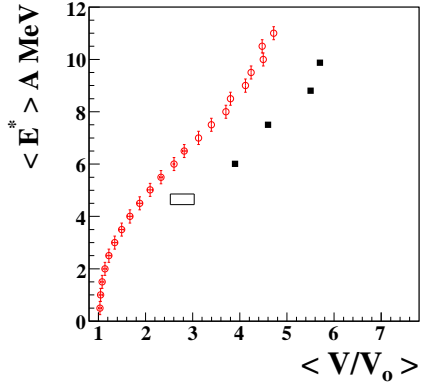}
\end{minipage}
\hspace{.05\textwidth}%
\begin{minipage}[c]{.45\textwidth}
\centering
\caption{Fragmentation position in the excitation energy-freeze-out volume
plane. The four full squares (QF sources) are taken from~\protect\cite{I66-Pia08}. 
The open rectangle gives the estimated position (with error bar) for QF
source at 25~AMeV, and the open circles those for QP sources.
From.~\protect\cite{I69-Bon08}}\label{fig4}
\end{minipage}
\end{figure}

To calibrate the freeze-out volumes for other sources,
we use the charge of the heaviest fragment $<Z_1^{(N)}>$ or the 
fragment multiplicity $<M_{frag}^{(N)}>$, normalized to the size of the 
source, as representative of the volume or density at break-up.
From the four points for QF sources and the additional constraint that
$Z_1^{(N)} = M_{frag}$=1 at $V/V_0$=1, we obtain two relations
$V/V_0 = f_1(Z_1^{(N)} )$ and  $V/V_0 = f_2(M_{frag}^{(N)})$, from which we
calculate the volumes for QF sources at 25~AMeV and for QP sources.
The results are plotted in fig.~\ref{fig4}, with error bars coming
from the difference between the two estimates using $f_1$ and $f_2$; note that 
error bars for the QP volumes are small up to 7~AMeV, and can not be 
estimated above, due to the fall of $<M_{frag}^{(N)}>$ at high energy (see
fig. 5 of~\cite{I69-Bon08}). So only $<(Z_1^{(N)}>)$ can be used
over the whole excitation energy range considered and the derived function
is the following:\\$V/V_0 = exp(2.47-4.47<(Z_1^{(N)}>)+0.86$. 

The volumes of QP sources are smaller than those of QF sources
(by about 20\% on the $E^*$ range 5-10~AMeV). This supports the explanation 
discussed previously starting from fluctuations of the charge of the 
heaviest fragment in a partition.

$Z_1$ also presents some specific dynamical properties. As shown 
in~\cite{I9-Mar97,I57-Tab05} for QF sources, 
its average kinetic energy is smaller than that of other fragments with the
same charge. The effect was observed whatever the fragment multiplicity for
Xe+Sn between 32 and 50 AMeV and for Gd+U at 36 AMeV. The
fragment-fragment correlation functions are also different when one
of the two fragments is $Z_1$.
This observation was
connected to the event topology at freeze-out, the heavier 
fragments being systematically closer to the centre of mass than the 
others.
\section{Heaviest fragment and order parameter}
\begin{figure}[th]
\centerline{\psfig{file=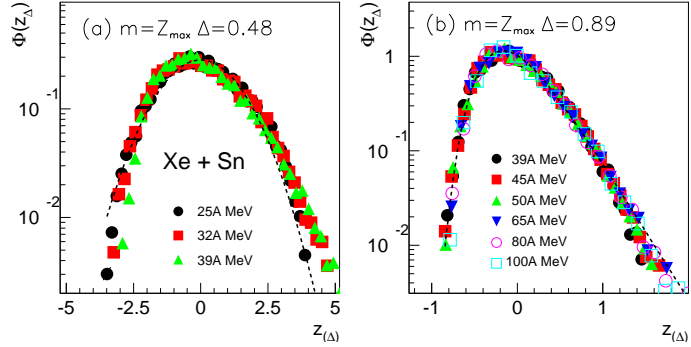,width=9cm}}
\vspace*{8pt}
\caption{ (a) $Z_{max}$ distributions for central Xe+Sn collisions at
25-39 AMeV bombarding energies, scaled according to $\Delta$ scaling
equation; the dashed curve is a best fit to scaled data using a Gaussian
distribution. (b) As (a) but for bombarding energies 39-100 AMeV:
the dashed curve is a best fit to scaled data using the Gumbel distribution.
 From.~\protect\cite{I51-Fra05}}\label{fig5}
\end{figure}

The recently developed theory of universal scaling laws of order-parameter
fluctuations 
provides methods to select order parameters.~\cite{Bot00,Bot02}
In this framework, 
universal $\Delta$ scaling laws of one of the order parameters, $m$, 
should be observed:\\ 
$ \langle m \rangle^{\Delta} P(m) = 
\phi ((m - \langle m \rangle )/ \langle m \rangle ^{\Delta}) $ \\
where $\langle m \rangle$ is the mean value of the distribution $P(m)$.
$\Delta$=1/2 corresponds to small fluctuations, 
$\sigma_m^2 \sim \langle m \rangle$, and
thus to an ordered phase. Conversely $\Delta$=1 occurs for the largest
fluctuations nature provides, $\sigma_m^2 \sim \langle m \rangle^2$, 
in a disordered phase. For models of cluster production
there are two possible order parameters:~\cite{Bot02} the 
fragment multiplicity in a fragmentation process or the size of the 
largest fragment in an aggregation process (clusters are
built up from smaller constituents).
The method was applied to central collision samples (symmetric systems with
total masses $\sim$73-400 at bombarding energies between
25 and 100 AMeV)~.\cite{I51-Fra05,Bot01}
The  total (charged products) or fragment multiplicity fluctuations do not
show any evolution over the 
whole data set. Conversely the relationship between the
mean value and the fluctuation of the size of
the largest fragment does change as a function of the bombarding energy:
$\Delta \sim$1/2 at low energy, and $\Delta \sim$1 for higher bombarding
energies. The form of the $Z_{max}$ distributions also evolves with
bombarding energy: it is nearly Gaussian in the $\Delta$=1/2 regime and
exhibits for $\Delta$=1 an asymmetric form with a near-exponential tail for
large values of the scaling variable (see fig.~\ref{fig5}). This
distribution is close to that of the modified Gumbel
distribution,~\cite{Gum58} the resemblance increasing with the total mass of
the system studied and being nearly perfect for the Au+Au data. The Gumbel
distribution is the equivalent of the Gaussian distribution in the case of
extreme values: it is obtained for an observable which is an extremum of a
large number of random, uncorrelated, microscopic variables.

Within the developed theory, this
behaviour indicates, for extensive systems, the transition from an ordered
phase to a disordered phase in the critical region, the fragments being
produced  following some aggregation scenario.
However simulations for finite systems have been performed
in the framework of the Ising model~\cite{Car02} which show that the
distribution of the heaviest fragment approximately obeys the $\Delta$=1 
scaling regime even at subcritical densities where no continuous 
transition takes place.
The observed behaviour was interpreted  as a finite size effect that
prevents the recognition of the order of a transition in a small system.
More recently the distribution of the heaviest fragment was analyzed within
the lattice gas model~\cite{Gul05} and it was shown that the most important
finite size effect comes from conservation laws, the distribution of the
order parameter being strongly deformed if a constraint is applied (mass
conservation) to an
observable that is closely correlated to the order parameter. Moreover the
observation of the $\Delta$=1 scaling regime was indeed observed in the
critical zone but was also confirmed at subcritical densities inside the
coexistence region. 
\section{Heaviest fragment and first order phase transition}
At a first-order phase transition, the distribution of the order parameter 
in a finite system presents a characteristic bimodal behavior in the 
canonical or grandcanonical ensemble.~\cite{Bin84,Cho01,Cho03,Gul03}
The bimodality comes from an anomalous convexity of the underlying 
microcanonical entropy.~\cite{Gros02} It physically corresponds to the 
simultaneous presence of two different classes of physical states for the same 
value of the control parameter, and can survive at the thermodynamic 
limit in a large class of physical systems subject to long-range 
interactions.~\cite{LNP02}

\begin{figure}[htb]
\begin{minipage}[c]{.45\textwidth}
\centering
\includegraphics[width=1.15\textwidth]
{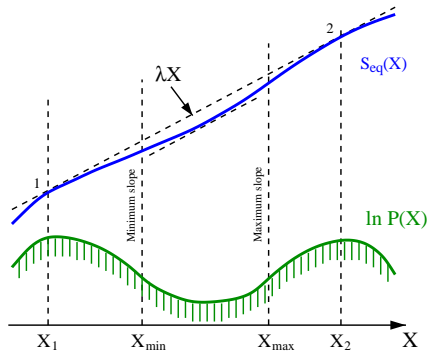}
\end{minipage}
\hspace{.05\textwidth}%
\begin{minipage}[c]{.45\textwidth}
\centering
\caption{
Canonical ensemble of finite systems. The bimodal equilibrium distribution
is given by $P(X)$$\sim$exp($S(X)$-$\lambda X$). The figure shows the case when
the Lagrange multiplier $\lambda$  is equal to the slope of the common tangent
(from~\protect\cite{Cho04}). \label{fig6}}
\end{minipage}
\end{figure}

Indeed if one considers a finite system  in contact with a
reservoir (canonical sampling), the value of the extensive variable (order
parameter) X may fluctuate as the system explores the phase space.
The entropy function F(X) is no more addititive due to the fact
that surfaces are not negligible for finite systems and the resulting
equilibrium entropy function has a local convexity. The Maxwell construction
is no longer valid.
The associated distribution at equilibrium
is $P(X)$$\sim$exp($S(X)$-$\lambda X$) where $\lambda$ is the corresponding
Lagrange multiplier. The distribution of $X$ 
acquires a bimodal
character (see fig.~\ref{fig6}).
In the case of nuclear multifragmentation, we have shown in the previous section 
that the size of the heaviest cluster produced in each collision 
event is an order parameter. A
difficulty comes however from the absence
of a true canonical sorting in the data. The statistical ensembles    
produced by selecting for example fused systems are
neither canonical nor microcanonical and should be better described in 
terms of the Gaussian ensemble,~\cite{Cha88} which gives a continuous
interpolation between canonical and microcanonical ensembles. Recently
a simple weighting of the probabilities associated to each excitation 
energy bin for quasi-projectile events was proposed to allow the comparison
with the canonical ensemble.~\cite{Gul07} That weighting 
procedure is used 
to allow a comparison
with canonical expectations for QP sources produced in Au + Au collisions
at incident energies from 60 to 100 AMeV. Then, a double 
saddle-point approximation is applied to extract from the measured data 
equivalent-canonical distributions.~\cite{Gul07}
\begin{figure}[htb]
\begin{minipage}[c]{.45\textwidth}
\centering
\includegraphics[width=1.\textwidth]
{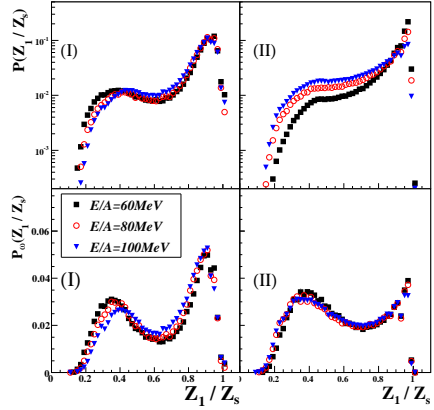}
\end{minipage}
\hspace{.05\textwidth}%
\begin{minipage}[c]{.45\textwidth}
\centering
\caption{Upper part: measured distribution of the charge of the largest fragment
normalized to the charge of the source detected in Au+Au collisions 
at three different bombarding energies. Lower part: weighted distributions 
obtained considering the same statistics for each excitation energy bin. 
The left (right) side shows distributions obtained with the data selection 
method (I) ((II)). From.~\protect\cite{I72-Bon09}} 
\label{fig7}
\end{minipage}
\end{figure}

In this incident energy regime, a part of the cross section corresponds 
to collisions with dynamical neck formation~\cite{DiT06}. 
We thus need to make sure that the observed change in the fragmentation
pattern~\cite{I61-Pic06} 
is not trivially due to a change in the size of the QP.
After a shape analysis in the center of mass frame~\cite{Cug83}, only events with a 
total forward detected charge larger than 80\% of the Au charge were 
considered (quasi-complete subevents).
Two different procedures aiming at selecting events with negligible neck contribution were adopted.
In the first one~\cite{I61-Pic06} (I) by eliminating  events where the entrance 
channel dynamics induces a forward emission, in the quasi-projectile frame, 
of the heaviest fragment $Z_1$.~\cite{I36-Col03}
For isotropically decaying QPs, this procedure does not bias the event sample
but only reduces the statistics.
In a second strategy (II) the reduction of the neck contribution is
obtained by keeping only ``compact''  events by 
imposing (i) an upper limit on the relative 
velocity among fragments, and (ii) a QP size constant
within 10\% (see~\cite{I69-Bon08} for details). 
In both cases fission events were removed.~\cite{I61-Pic06}

The results obtained with the two different selection methods are given in 
fig.~\ref{fig7}. To take into account the small variations
of the source size, the charge of the heaviest fragment $Z_1$ has been normalized
to the source size. 
After the weighting procedure (lower part of the figure), a bimodal behavior of the largest
fragment charge clearly emerges in both cases.
In particular in the case of selection (II), we can see that the weight of the low $Z_1$ component, 
associated to more fragmented configurations and higher deposited energy, 
increases with the bombarding energy before the weighting procedure (upper
part of the figure). This difference
completely disappears when data are 
weighted, showing the validity of the phase-space dominance hypothesis.

Those weighted experimental distributions can be fitted with an
analytic function (see~\cite{I72-Bon09} for more details).
From the obtained parameter values one can estimate the
latent heat of the transition of
the hot heavy nuclei studied (Z$\sim$70) as 
$\Delta E=8.1 (\pm0.4)_{stat} (+1.2 -0.9)_{syst}$~AMeV.
Statistical error was
derived from experimental statistics and systematic errors
from the comparison between the two different QP selections.


\begin{thebibliography}{10}
\expandafter\ifx\csname url\endcsname\relax
  \def\url#1{\texttt{#1}}\fi
\expandafter\ifx\csname urlprefix\endcsname\relax\def\urlprefix{URL }\fi
\providecommand{\eprint}[2][]{\url{#2}}

\bibitem{Boh36}
N.~Bohr, \emph{Nature} \textbf{137} (1936) 351.

\bibitem{I46-Bor02}
B.~Borderie, \emph{J. Phys. G: Nucl. Part. Phys.} \textbf{28} (2002) R217.

\bibitem{WCI06}
P.~Chomaz et~al. (eds.) vol.~30 of \emph{Eur. Phys. J. A}, Springer, 2006.

\bibitem{Bor08}
B.~Borderie and M. F. Rivet, \emph{Prog. Part. Nucl. Phys.} \textbf{61} (2008) 551.

\bibitem{Sou06}
R.~T. {de Souza} et~al., P.~Chomaz et~al. (eds.) Dynamics and Thermodynamics
  with nuclear degrees of freedom, Springer, 2006, vol.~30 of \emph{Eur. Phys.
  J. A}, 275--291.

\bibitem{I3-Pou95}
J.~Pouthas et~al., \emph{Nucl. Instr. and Meth. in Phys. Res.} \textbf{A 357}
  (1995) 418.

\bibitem{I14-Tab99}
G.~T\u{a}b\u{a}caru et~al. (INDRA Collaboration), \emph{Nucl. Instr. and Meth.
  in Phys. Res.} \textbf{A 428} (1999) 379.
  
\bibitem{I33-Par02}
M.~Pârlog et~al. (INDRA Collaboration), \emph{Nucl. Instr. and Meth. in Phys.
  Res.} \textbf{A 482} (2002) 674.

\bibitem{I34-Par02}
M.~Pârlog et~al. (INDRA Collaboration), \emph{Nucl. Instr. and Meth. in Phys.
  Res.} \textbf{A 482} (2002) 693.

\bibitem{I69-Bon08}
E.~Bonnet et~al. (INDRA and ALADIN Collaborations), \emph{Nucl. Phys.}
  \textbf{A 816} (2009) 1.

\bibitem{I12-Riv98}
M.~F. Rivet et~al. (INDRA Collaboration), \emph{Phys.\ Lett.} \textbf{B 430}
  (1998) 217.

\bibitem{Gul06}
F.~Gulminelli et~al., P.~Chomaz et~al. (eds.) Dynamics and Thermodynamics with
  nuclear degrees of freedom, Springer, 2006, vol.~30 of \emph{Eur. Phys. J.
  A}, 253--262.

\bibitem{I63-NLN07}
N.~{Le Neindre} et~al. (INDRA and ALADIN collaborations), \emph{Nucl. Phys.}
  \textbf{A 795} (2007) 47.

\bibitem{MDA04}
M.~D'Agostino et~al., \emph{Nucl. Phys.} \textbf{A 734} (2004) 512.

\bibitem{T41Bon06}
E.~Bonnet, thèse de doctorat, Université Paris-XI Orsay (2006),
  {http://tel.archives-ouvertes.fr/tel-00121736}.

\bibitem{I58-Pia05}
S.~Piantelli et~al. (INDRA Collaboration), \emph{Phys. Lett.} \textbf{B 627}
  (2005) 18.

\bibitem{I66-Pia08}
S.~Piantelli et~al. (INDRA Collaboration), \emph{Nucl. Phys.} \textbf{A 809}
  (2008) 111.

\bibitem{Koo87}
S. E. Koonin and J. Randrup, \emph{Nucl. Phys.} \textbf{A 474} (1987) 173.

\bibitem{I39-Hud03}
S.~Hudan et~al. (INDRA Collaboration), \emph{Phys. Rev.} \textbf{C 67} (2003)
  064613.

\bibitem{Kim92}
Y.~D. Kim et~al., \emph{Phys. Rev.} \textbf{C 45} (1992) 338.

\bibitem{Bow95}
D.~R. Bowman et~al., \emph{Phys. Rev.} \textbf{C 52} (1995) 818.

\bibitem{Gro97}
D.~H.~E. Gross, \emph{Phys. Rep.} \textbf{279} (1997) 119.

\bibitem{I57-Tab05}
G.~T\u{a}b\u{a}caru et~al. (INDRA Collaboration), \emph{Nucl. Phys.} \textbf{A
  764} (2006) 371.

\bibitem{I9-Mar97}
N. Marie et~al. (INDRA Collaboration), \emph{Phys. Lett.} \textbf{B 391}
 (1997) 15.

\bibitem{Bot00}
R.~Botet et~al., \emph{Phys. Rev.} \textbf{E 62} (2000) 1825.

\bibitem{Bot02}
R.~Botet et~al., Universal fluctuations, World Scientific, 2002, vol.~65 of
  \emph{World scientific Lecture Notes in Physics}.

\bibitem{I51-Fra05}
J.~D. Frankland et~al. (INDRA and ALADIN collaborations), \emph{Phys. Rev.}
  \textbf{C 71} (2005) 034607.

\bibitem{Bot01}
R.~Botet et~al., \emph{Phys. Rev. Lett.} \textbf{86} (2001) 3514.

\bibitem{Gum58}
E.~J. Gumbel, Statistics of extremes, Columbia University Press, 1958.

\bibitem{Car02}
J.~M. Carmona et~al., \emph{Phys. Lett.} \textbf{B 531} (2002) 71.

\bibitem{Gul05}
F.~Gulminelli et~al., \emph{Phys. Rev.} \textbf{C 71} (2005) 054607.

\bibitem{Bin84}
K.~Binder et~al., \emph{Phys. Rev.} \textbf{B 30} (1984) 1477.

\bibitem{Cho01}
P.~Chomaz et~al., \emph{Phys. Rev.} \textbf{E 64} (2001) 046114.

\bibitem{Cho03}
P.~Chomaz et~al., \emph{Physica} \textbf{A 330} (2003) 451.

\bibitem{Gul03}
F.~Gulminelli, \emph{Ann. Phys. Fr.} \textbf{29} (2004) N$^{o}$ 6.

\bibitem{Gros02}
D.~H.~E. Gross, Microcanonical thermodynamics, World Scientific, 2002, vol.~66
  of \emph{World scientific Lecture Notes in Physics}.

\bibitem{LNP02}
T.~Dauxois et~al. (eds.) vol. 602 of \emph{Lecture Notes in Physics},
  Springer-Verlag, Heidelberg, 2002.

\bibitem{Cho04}
P.~Chomaz et~al., \emph{Phys. Rep.} \textbf{389} (2004) 263.

\bibitem{Cha88}
M.~Challa et~al., \emph{Phys. Rev. Lett.} \textbf{60} (1988) 77.

\bibitem{Gul07}
F.~Gulminelli, \emph{Nucl. Phys.} \textbf{A 791} (2007) 165.

\bibitem{I72-Bon09}
E.Bonnet et~al. (INDRA Collaboration), \emph{Phys. Rev. Lett.} \textbf{103}
  (2009) 072701.

\bibitem{DiT06}
M.~D. Toro et~al., P.~Chomaz et~al. (eds.) Dynamics and Thermodynamics with
  nuclear degrees of freedom, Springer, 2006, vol.~30 of \emph{Eur. Phys. J.
  A}, 65--70.

\bibitem{I61-Pic06}
M.~Pichon et~al. (INDRA and ALADIN collaborations), \emph{Nucl. Phys.}
  \textbf{A 779} (2006) 267.

\bibitem{Cug83}
D.~Cugnon et~al., \emph{Nucl. Phys.} \textbf{A 397} (1983) 519.

\bibitem{I36-Col03}
J.~Colin et~al. (INDRA Collaboration), \emph{Phys. Rev.} \textbf{C 67} (2003)
  064603.

\end{thebibliography}

\end{document}